\documentclass{ws-procs975x65}

\begin{document}

\title{Twisted reduction in large $N$ QCD with adjoint Wilson fermions}

\author{Antonio Gonz\'alez-Arroyo$^{ab}$}

\address{{$^a$}Instituto de F\'{\i}sica Te\'orica UAM/CSIC\\
{$^b$}Departamento de F\'{\i}sica Te\'orica, C-15\\
Universidad Aut\'onoma de Madrid, E-28049--Madrid, Spain\\
E-mail: antonio.gonzalez-arroyo@uam.es}

\author{Masanori Okawa$^c$}

\address{{$^c$}Graduate School of Science, Hiroshima University\\
Higashi-Hiroshima, Hiroshima 739-8526, Japan\\
E-mail: okawa@sci.hiroshima-u.ac.jp}

\begin{abstract}
The twisted reduced model of large $N$ QCD with two adjoint Wilson fermions 
is studied numerically using the Hybrid Monte Carlo method. 
This is the one-site model, whose large $N$ limit (large volume limit) 
is expected to be conformal or nearly conformal. 
The string tension calculated at $N$=289 approaches zero 
as we decrease quark mass and the preliminary value of the mass anomalous dimension 
$\gamma_*$ is close to one if we assume that the theory is governed by an infrared 
fixed point. We also discuss the twisted reduced model with single adjoint Wilson fermion.
The string tension remains finite as the quark mass decreases to zero, supporting 
that this is the confining theory.
\end{abstract}

\keywords{large $N$ QCD, twisted space-time reduction, adjoint fermion}

\bodymatter

\section{Introduction}
The standard model of the elementary particle has set the foundation to SU(N) gauge theories. 
Although SU(N) gauge theories generally have very complicated structures, significant 
simplifications could occur in the large $N$ limit.  In fact, Eguchi and Kawai considered the space-time 
reduced model of the lattice gauge theory having only one space-time point\cite{EK}.  
This model is now called 
the Eguchi-Kawai model (EK-model).  The EK model possesses the Z(N) symmetry.  If this symmetry is not spontaneously broken, the lattice gauge theory and the EK model are equivalent in the large $N$ limit.
This means that space-time degrees of freedom of the lattice gauge theory could be  encoded into those of internal 
SU(N) group.  
Soon after the proposal of Eguchi and Kawai, however, the Z(N)
symmetry was shown to be spontaneously broken
in the weak coupling limit, thus invalidating the EK model\cite{QEK}.
To circumvent this difficulty, the present authors have 
proposed to consider the one-site model with twisted boundary conditions (the twisted Eguch-Kawai (TEK) model), 
thus avoiding the Z(N) symmetry being spontaneously broken\cite{TEK,TEK2}.  Recently, gauge theories 
coupled with fermions  in the adjoint representation received much attention in the large $N$ limit in
connection, for example, to the AdS/CFT correspondence.  The idea of the twisted space-time reduction
can also be applied to these models.  The purpose of this talk is to
present our recent progress in this
direction. In the next section, we calculate the continuum string tension of the SU(N) pure gauge theory in 
the large $N$  limit using the TEK model.  In sect. 3, large $N$ QCD with two adjoint Wilson fermions
is studied 
with the twisted space-time reduction, and sect. 4 is devoted to the large $N$ QCD with single adjoint 
Wilson fermion.  Conclusions are  given in sect. 5. 
 
\vspace{-0.3cm} 
 
\section{The TEK model}

We consider the SU(N) group with $N=L^2$, $L$ being some positive integer.  Then the action of the  
TEK model is obtained from the action of the SU(N) lattice gauge theory

\begin{equation}
 S= -b N \sum_n \sum_{\mu \ne \nu =1}^4 {\rm Tr} \left[ U_{n,\mu} U_{n+\mu,\nu} U_{n+\nu,\mu}^\dagger U_{n,\nu}^\dagger \right]
\end{equation}

\noindent
by neglecting the space-time dependence of the link variable 
$U_{n,\mu} \rightarrow U_\mu$
and multiplying the twist tensor $Z_{\mu\nu}$ in front of the plaquette action as\cite{TEK}

\begin{equation}
S_{TEK}=-b N \sum_{\mu \ne \nu =1}^4 {\rm Tr} \left[ Z_{\mu\nu} U_\mu U_\nu U_\mu^\dagger U_\nu^\dagger \right] 
\end{equation} 

\noindent
As a result, there remain only four SU(N) matrices $U_{\mu}$. The inverse 't Hooft coupling
is labelled $b=1/g^2N$.  The elements of the twist tensor belong to Z(L) and the explicit form is given by

\begin{equation}
 Z_{\mu\nu} = \exp \left( k {2\pi i \over L} \right), \ \ \ Z_{\nu\mu}=Z_{\mu\nu}^*, \ \ \ \mu>\nu 
\end{equation}

\noindent
The action $S_{TEK}$ is invariant under the Z(L) transformation $U_{\mu} \rightarrow z U_{\mu}$, 
$z \in$ Z(L).  If this symmetry is not spontaneously broken, the TEK model and the corresponding lattice gauge theory are equivalent in the large $N$ limit.  $k$ and $L$ should be coprime, and a general prescription for choosing $k$ and $L$ to preserve the Z(L) symmetry has been given in ref. [\refcite{TEK2}].  In this talk, we choose $L$=29 and $k$=11 for the TEK model\cite{TEK3,TEK4}.  For finite $L$, the TEK model is closely related to the lattice 
gauge theory on a finite space-time volume $L^4$, and we are able to calculate Wilson loops up to 
size $14 \times 14$ for $L$=29. 
In the TEK model, the Wilson loop $W(R,T)$ with size $R \times T$ is obtained from four link variables 
$U_\mu$ as 

\begin{equation}
 W(R,T) =  Z_{\mu\nu}^{RT} < U_\mu^R U_\nu^T U_\mu^{R \dagger} U_\nu^{T \dagger} >.
\label{WILSON} 
\end{equation}

\noindent
Then the string tension $\sigma$ is extracted from the large distance behavior of the Creutz ratio

\begin{equation}
 \chi(R',T')=-\log{W(R'+0.5,T'+0.5)W(R'-0.5,T'-0.5) \over W(R'+0.5,T'-0.5)W(R'-0.5,T'+0.5) }
\end{equation}

\noindent 
as
\vspace{-0.2cm}

\begin{equation}
 \chi(R',R')=\sigma + {2\gamma \over R'^2} + {\eta \over R'^4} 
\end{equation}

\noindent 
with half-integer $R'$ and $T'$.  We calculate the continuum string tension by extrapolating 
the TEK data at six values of  $b$.  For comparison, we also calculate the continuum string
tension of the usual SU(N) lattice gauge theory with $N=3, 4 ,5 ,6, 8$ on a $V=32^4$ lattice.
Results are summarized in Fig. 1.  The $\times$ point is $\Lambda_{\overline {MS}}/\sqrt{\sigma}$
for the TEK model with $N$=841.  The $+$ symbols are results for the lattice gauge theory, and dotted 
line is a linear fit of these data as a function of $1/N^2$.
The large $N$ extrapolated value agrees remarkably well with the result of the TEK model, 
demonstrating the correctness of the twisted space-time reduction idea.
      
\begin{figure}[t]
\begin{center}
\psfig{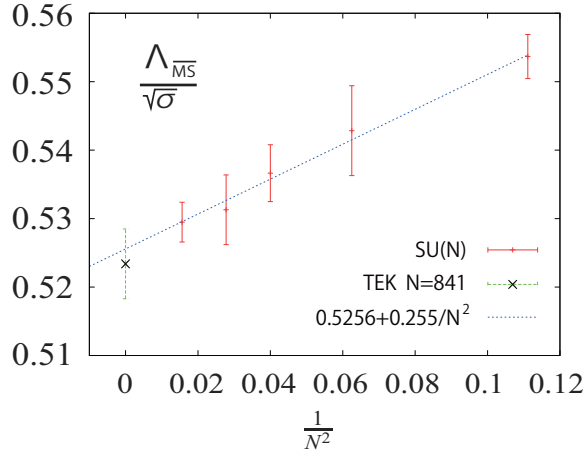}
\end{center}
\caption{$N$ dependence of the continuum string tension.}
\label{aba:fig1}
\end{figure}

\vspace{-0.3cm} 

\section{Large $N$ QCD with two adjoint fermions}

SU(N) gauge theories with two adjoint fermions are thought to be conformal or nearly conformal 
for any value of $N$, since the first two coefficients of the beta function expressed in term of
't Hooft coupling are independent of $N$, namely, $b_0=(4N_f-11)/24\pi^2$ and 
$b_1=(16N_f-17)/192\pi^4$.  For $N_f=2$, $b_0<0$ and $b_1>0$, then we naturally expect that 
there is a infrared fixed point for finite value of 't Hooft coupling.  In fact, for $N$=2
(minimal walking technicolor),
there are now many lattice simulations indicating that the theory is conformal at vanishing 
fermion mass\cite{DD}. The mass anomalous dimension, however, is shown to be rather small 
$\gamma_* \sim 0.3$, to explain the large values of the observed quark
masses, within the walking technicolor scenario.

We can also consider the twisted space-time reduced model of large $N$ QCD with two adjoint fermions.
\footnote{For previous works of the space-time reduced model of large $N$ QCD with adjoint fermions, 
see ref. [\refcite{KUY,AHUY,BKS}]. }
The action reads

\begin{equation}
S=-b N \sum_{\mu \ne \nu =1}^4 {\rm Tr} \left[ Z_{\mu\nu} U_\mu U_\nu U_\mu^\dagger U_\nu^\dagger \right] 
+ \sum_{j=1}^{N_f} {\bar \Psi}_j D_W \Psi_j
\end{equation} 

\noindent
with $N_f$=2 and the Wilson-Dirac operator $D_W$ given by

\begin{equation}
D_W=1-\kappa \sum_{\mu=1}^4 \left[ (1-\gamma_\mu) U_\mu^{adj} + (1+\gamma_\mu) U_\mu^{\dagger adj} \right]
\end{equation} 

\noindent
$\Psi_j$ is the fermion matrix in the color $(N,{\bar N})$ representation.  Thus the link variable in the 
adjoint representation $U_\mu^{adj}$ actually acts on $\Psi_j$ as $U_\mu^{adj} \Psi_j = U_\mu \Psi_j 
U_\mu^\dagger$. The  hopping parameter  of the fermion field  $\kappa$
is related to the bare fermion mass as
$m_f=(1/2)( 1/\kappa - 1/\kappa_c)$
with $\kappa_c$ the critical hopping parameter.

We calculate the string tension  at two values of $b$, 0.35 and 0.36, and for 
various values of $\kappa$. We are using  $N=289=17^2$ and $k=5$,
which should compare with ordinary  lattice theory with $V=17^4$\cite{MO}. 
For $N_f=2$, we can use the standard hybrid Monte Carlo method.  Simulations have been done on
Hitachi SR16000 supercomputer at KEK having peek speed of 980 GFlops per node.  Thanks to the Hitachi system 
engineers, our code is highly optimized for the  SR16000. The sustained speed of our code being 600 GFlops. 

\begin{figure}[t]
\begin{center}
\psfig{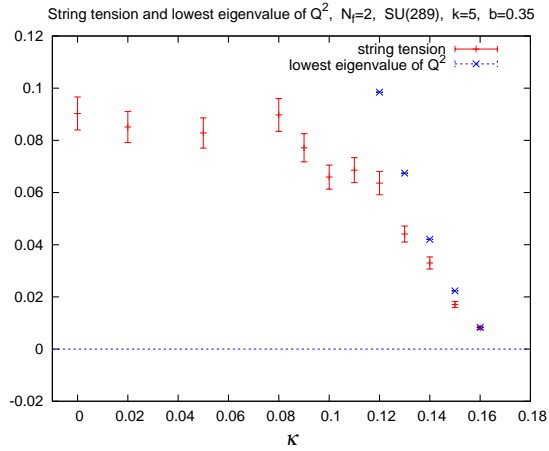}
\end{center}
\caption{String tension and lowest eigenvalue of $Q^2$ at $b$=0.35 with $N_f=2$.}
\label{aba:fig2}
\end{figure} 
 
In Fig. 2, we show the string tension $\sigma$ as a function of $\kappa$ at $b$=0.35 with $+$ symbols. 
The value of $\sigma$ at $\kappa$=0 is obtained from the TEK model without fermions.  We clearly see that as 
$\kappa$ increases the string tension rapidly decreases and seems to vanish at $\kappa \sim 0.17$.
So far, we have not calculated any hadronic spectrum.  It is quite straightforward, however, to calculate 
the lowest eigenvalue of the positive hermitian Wilson-Dirac operator $Q^2=(D_W \gamma_5)^2$, which should be 
related to the physical fermion mass square.  In Fig. 2, we also show the lowest eigenvalue of $Q^2$ as $\times$ 
symbols.  The string tension and the lowest eigenvalue of $Q^2$ vanish simultaneously, which strongly supports 
that the string tension is zero at critical $\kappa_c$ where the fermion mass vanishes.

We can fit both the string tension $\sigma$ and the lowest eigenvalue of $Q^2$ with the same fitting form 
$a ( 1/\kappa - 1/\kappa_c )^b$.  Requiring that both quantities vanish at the same critical $\kappa_c$,
we have $\kappa_c=0.1694(7)$ and, for the string tension, $b=1.10(4)$.  If the massless theory is governed by an 
infrared fixed point, $b$ is related to the mass anomalous dimension $\gamma_*$ as $b=2/(1+\gamma_*)$, then 
we have $\gamma_*=0.81(8)$. 

\begin{figure}[t]
\begin{center}
\psfig{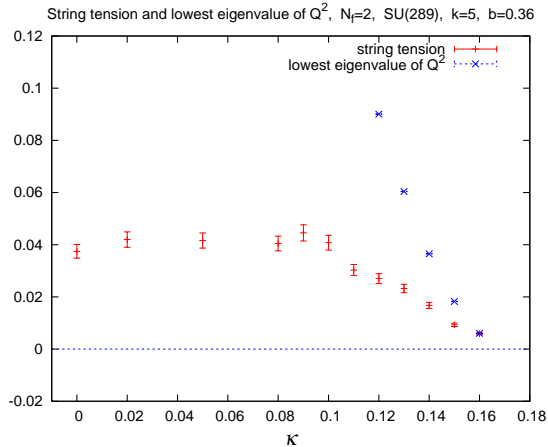}
\end{center}
\caption{String tension and lowest eigenvalue of $Q^2$ at $b$=0.36 with $N_f=2$.}
\label{aba:fig2}
\end{figure} 

We have repeated the analysis at $b$=0.36, a little bit closer to the continuum limit.  Results are
shown in Fig. 3.  It seems that the data at $\kappa=0.16$ are largely affected by the finite -size effects.
Excluding the data at $\kappa=0.16$ from our analysis, we have $\kappa_c=0.168(1)$, $b=1.33(4)$, and thus 
$\gamma_*=1.17(21)$.  Although our data are still preliminary, large $N$ QCD with two adjoint fermions seems to 
be a conformal field theory with large mass anomalous dimension $\gamma_* \sim 1$.  

\vspace{-0.3cm} 

\section{Large $N$ QCD with single adjoint fermion}

In the large $N$ limit, $N_f=1$ adjoint fermion is equivalent to the $N_f=2$ fundamental fermions in the
rank two anti-symmetric representation (the orientifold equivalence\cite{KUY}).  For $N=3$, the latter 
theory is just the two flavor QCD.  This means that the large $N$ QCD with single adjoint fermion corresponds to the Corrigan-Ramond large $N$ limit\cite{CR} of realistic QCD, and thus it is quite interesting to analyze this theory from the phenomenological point of view.

\begin{figure}[t]
\begin{center}
\psfig{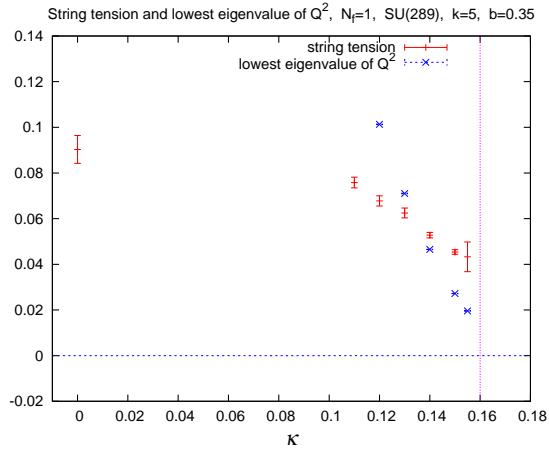}
\end{center}
\caption{String tension and lowest eigenvalue of $Q^2$ at $b$=0.35 with $N_f=1$.}
\label{aba:fig2}
\end{figure}

\begin{figure}[t]
\begin{center}
\psfig{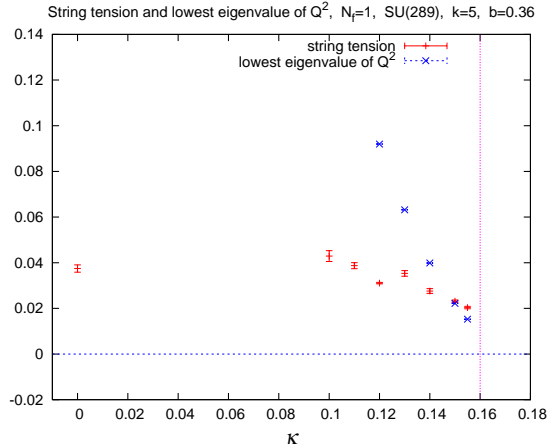}
\end{center}
\caption{String tension and lowest eigenvalue of $Q^2$ at $b$=0.36 with $N_f=1$.}
\label{aba:fig2}
\end{figure}     

We use the rational hybrid Monte Carlo method to simulate the $N_f=1$ theory.  
In Fig. 4, we plot the string tension $\sigma$ as a function of $\kappa$ at $b$=0.35 with $+$ symbols.  
Also shown is the lowest eigenvalue of $Q^2$ with $\times$ symbols.  The lowest eigenvalue of $Q^2$ 
seems to vanish around $\kappa \sim 0.165$, while  the string tension
seems to remain finite at this point, indicating 
that the large $N$ QCD with single adjoint fermion is a confining theory.  This is quite natural 
since our theory should be related to the two flavor $N=3$ QCD.  
We found that, for $\kappa \ge$ 0.16, we cannot make simulations since the CG iteration does not 
converge, which is also the common phenomena encountered in the simulation of  QCD.

In Fig. 5, we show the results at $b$=0.36.  The absolute values of the string tension are smaller than 
those of $b=0.35$ since we are approaching to the continuum limit.  The string tension, however, 
remains finite  around $\kappa \sim 0.165$ as in Fig. 4.   We also confirm that the CG iteration does
not converge for $\kappa \ge 0.16$.  Hence,  both Figs. 4 and 5 indicate that the continuum theory of the
large $N$ QCD with single adjoint fermion is a confining theory. 

\vspace{-0.3cm} 

\section{Conclusion}

We have demonstrated that the twisted reduction works quite well for the description of large $N$ QCD.  
For pure gauge theory, we have succeeded in calculating the continuum string tension using TEK model.
Although our calculations with dynamical fermions are still preliminary, we have also shown that large $N$ QCD with two adjoint fermions is a conformal theory with large mass anomalous 
dimension $\gamma_* \sim 1$, and that the theory with single adjoint fermion is a confining theory.
Since the rational hybrid Monte Carlo method we used for the simulation with one flavor can directly 
applicable to any number of flavor, we are planning to extend our analysis to $N_f=1/2$ super-symmetric 
theory and also  $N_f=3/2$ and $5/2$ theories if they are conformal or confining theories.   

\vspace{-0.3cm} 

\noindent
\section*{Acknowledgments}

A.G-A is supported by grants FPA2009-08785, FPA2009-09017, CSD2007-00042, HEPHACOS
S2009/ESP-1473 and ITN PITN-GA-2009-238353. M. O. is supported by the Japanese
MEXT grant No 23540310. The main calculations have been done on SR16000 computer at KEK supported
by the Large Scale Simulation Prog. No.12-01 (FY2011-12).  The calculations have also been done on 
SR16000 computer at YITP in Kyoto University and on the INSAM cluster system at Hiroshima University.

\end{document}